\documentclass[]{spie}  

\usepackage{amsmath,amsfonts,amssymb}
\usepackage{graphicx}
\usepackage[colorlinks=true, allcolors=blue]{hyperref}

\title{Detecting and characterizing close-in exoplanets with Vortex~Fiber~Nulling}

\author[a,$\dagger$]{Daniel~Echeverri}
\author[b]{Garreth~Ruane}
\author[a]{Benjamin~Calvin}
\author[a]{Nemanja~Jovanovic}
\author[c]{Jacques-Robert~Delorme}
\author[a]{Jason~Wang}
\author[a]{Maxwell~Millar-Blanchaer}
\author[a,b]{Dimitri~Mawet}
\author[b]{Eugene~Serabyn}
\author[b]{J.~Kent~Wallace}
\author[b]{Stefan~Martin}
\affil[a]{Department of Astronomy, California Institute of Technology, 1200 E. California Blvd.\\Pasadena, CA 91125, USA}
\affil[b]{Jet Propulsion Laboratory, California Institute of Technology, 4800 Oak Grove Dr.
\\Pasadena, CA 91109, USA}
\affil[c]{W. M. Keck Observatory, 65-1120 Mamalahoa Highway, Kamuela, HI 96743, USA.}

\authorinfo{$^\dagger$Send correspondence to dechever@caltech.edu}

\graphicspath{{Figures/}}
\begin{document} 
\maketitle

\begin{abstract}
Vortex Fiber Nulling (VFN) is an interferometric method for suppressing starlight to detect and spectroscopically characterize exoplanets. It relies on a vortex phase mask and single-mode fiber to reject starlight while simultaneously coupling up to 20\% of the planet light at separations of $\lesssim1\lambda/D$, thereby enabling spectroscopic characterization of a large population of RV and transit-detected planets, among others, that are inaccessible to conventional coronagraphs. VFN has been demonstrated in the lab at visible wavelengths and here we present the latest results of these experiments. This includes polychromatic nulls of $5\times10^{-4}$ in 10\% bandwidth light centered around 790~nm. An upgraded testbed has been designed and is being built in the lab now; we also present a status update on that work here. Finally, we present preliminary K-band (2~$\mu$m) fiber nulling results with the infrared mask that will be used on-sky as part of a VFN mode for the Keck Planet Imager and Characterizer Instrument in 2021.  
\end{abstract}

\keywords{Exoplanets, Interferometry, Instrumentation, High Dispersion Coronagraphy, Keck Telescope, Fiber Nulling}

\section{INTRODUCTION}
\label{sec:intro}  

Since its initial introduction in 1978 by Ronald N. Bracewell\cite{Bracewell1978}, nulling interferometry has remained one of the most promising techniques for detecting exoplanets at small angular separations. Bracewell's original concept, known as a Bracewell Interferometer, combined two symmetric telescope apertures to create a destructive interference fringe over the target star. The surrounding bright fringes were then rotated to modulate the planet signal while keeping the star centered on the dark fringe, or ``null", so that its signal remained constantly suppressed\cite{Bracewell1979}. In this way, the weak planet light could be extracted from behind the star's overwhelming glare. From that concept, nulling interferometry developed and advanced until eventually it became the underlying principle behind two major space telescope missions, NASA's Terrestrial Planet Finder Interferomter (TPF-I)\cite{Lawson2007_TPFI} and ESA's Darwin\cite{Kaltenegger2005_Darwin}. Both missions were ultimately cancelled but nulling interferometry continued to evolve over time and has seen other recent on-sky successes. The Palomar Fiber Nuller (PFN), for example, is a derivative of the Bracewell Interferometer that uses a pupil mask to make two identical sub-apertures on a single telescope instead of relying on two separate telescopes\cite{Haguenauer2006_PFN,Mennesson2006_PFN,Martin2008_PFN,Serabyn2010_PFN}. PFN also improves the stellar suppression by focusing the interference fringes onto a single-mode fiber (SMF). These two improvements on traditional nulling interferometry have allowed PFN to demonstrate on-sky null depths of $\sim10^{-3}$ in the past\cite{Mennesson2010_PFN,Mennesson2011_PFN} and more recently of $\sim10^{-4}$ with the detection of a faint companion well within the diffraction limit of the Palomar Telescope\cite{Serabyn2019_PFN}.

Building on this heritage, Vortex Fiber Nulling (VFN) is a new fiber nulling technique introduced by our team in 2018\cite{Ruane2018_VFN}. It is very similar to the PFN in that it relies on a pupil mask and SMF but instead of creating two sub-apertures, it uses a vortex mask\cite{Swartzlander2001} that allows it to utilize the full telescope aperture. This improves the total system throughput and also simplifies the optical layout since it removes the need for baseline rotation. VFN is also insensitive to the aperture shape, segmentation, or obstruction\cite{Ruane2019SPIE}. This simple design makes VFN easily deployable in any instrument with an available upstream pupil plane or focal plane for the vortex and the ability to inject light into a fiber. By combining VFN with a high resolution spectrograph, it is possible to spectrally detect and characterize exoplanets that would be very difficult to reach with conventional coronagraphs. As such, VFN's small inner-working-angle makes it ideal for spectroscopic followups of radial velocity (RV)- and transit-detected exoplanets such as Proxima~Centauri~b\cite{Escude2016_ProxCen} and Ross~128~b\cite{Bonfils2017_Ross128}.

In this paper we briefly summarize previous VFN laboratory demonstrations before going into more recent results of monochromatic and polychromatic experiments at visible wavelengths. We also present the final design for a new VFN testbed at Caltech as well as an update on its build and alignment status. Finally, we present preliminary results in K-band light ($2~\mu m$) with the vortex mask that will be used for an on-sky demonstration of VFN in the Keck Planet Imager and Characterizer (KPIC) instrument. 

\section{VFN CONCEPT}
\label{sec:VFN_Concept}
\begin{figure}[t!]
    \centering
    \includegraphics[width=\linewidth]{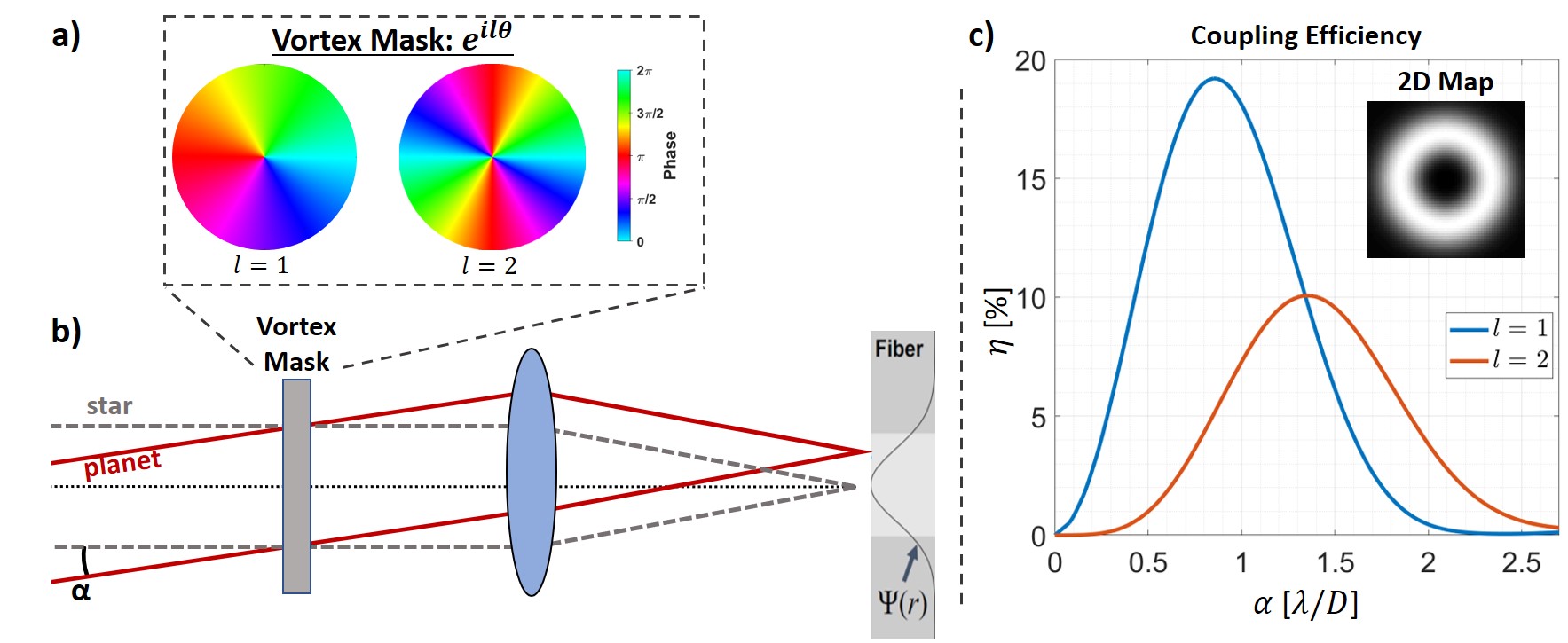}
    \caption{(a) The azimuthally varying phase pattern introduced by a charge $l=1$ and $l=2$ vortex mask. (b) Diagram of a VFN system with the vortex mask in the pupil plane and SMF in the image plane. The fundamental mode of the fiber can be accurately approximated by a gaussian and is denoted $\Psi(r)$. The star is aligned with the SMF such that the target planet lands at an off-axis angle $\alpha$. (c) Coupling efficiency, $\eta$, of a point source versus its angular separation from the optical axis, $\alpha$, for a charge $l=1$ (blue) and $l=2$ (orange) VFN system. The inset shows the coupling efficiency for all points in a field of view centered on the star/fiber.}
    \label{fig:MotherFig}
\end{figure}

The basic elements needed for Vortex Fiber Nulling are a vortex mask and a single mode fiber. A vortex mask is an optic that imparts an azimuthally varying phase ramp of the form $\exp(il\theta)$ on the light, where $l$ is an integer known as the charge and $\theta$ is the azimuthal angle. The charge thus defines the number of times the phase wraps from 0 to 2$\pi$ over the unit circle. A charge 1 vortex, for example, goes from 0 to 2$\pi$ once while a charge 2 vortex completes the cycle twice as shown in Fig.~\ref{fig:MotherFig}a. VFN can be implemented with the vortex mask in either a focal or pupil plane while achieving similar performance as explained by Ruane et al.\cite{Ruane2019SPIE}. Nevertheless, for simplicity and to match the layout we have used in the lab and plan to use on sky, this paper will only deal with the pupil-plane version of VFN as shown in Fig.~\ref{fig:MotherFig}b. By placing the vortex in the pupil plane, all point sources in the field are imparted with the same azimuthal phase ramp of the vortex. 

The SMF is then placed in an image plane downstream of the vortex and centered on the optical axis so that when the star is aligned to the fiber, its phase is symmetrically varying on the fiber core. The coupling efficiency, $\eta$, or the fraction of light that couples into the SMF, from any source in the field can be computed using the overlap integral:

\begin{equation}
    \eta(\alpha)=\left|\int E(\mathbf{r};\alpha) \Psi(r)dA\right|^2,
    \label{eqn:couplingeff}
\end{equation}
where $E(\mathbf{r};\alpha)$ is the field at the entrance to the SMF, $\Psi(r)$ is the fundamental mode of the SMF, $\mathbf{r}=(r,\theta)$ are the coordinates in the fiber-tip plane, and $\alpha$ is the angular offset with respect to the optical axis (see Fig.~\ref{fig:MotherFig}b). The fundamental mode of a SMF can be accurately approximated as a Gaussian, thereby providing the form for $\Psi(r)$. $E(\mathbf{r};\alpha)$ depends on the position of the given source but for the on-axis star, its phase becomes symmetric on the fiber mode. This causes the integral to vanish and the star is thus rejected from coupling into the fiber. The electric field from an off-axis point, such as a planet, is not symmetric on the fiber tip though. The integral is thus non-zero and the planet light couples into the fiber. Figure~\ref{fig:MotherFig}c shows the coupling efficiency as a function of the angular separation, $\alpha$, for any point in the field as evaluated using Eq.~\ref{eqn:couplingeff}.

Each vortex charge has its own coupling efficiency profile; Fig.~\ref{fig:MotherFig}c shows the charge 1 and charge 2 cases. The VFN planet coupling can be read directly from this plot. For example, if the planet is located at $0.9\lambda/D$, 19\% of its light couples into the fiber when a charge 1 vortex mask is used while only 6\% couples in for charge 2. Charge 1 is generally favored for planets within the diffraction limit while charge 2 has higher throughput for larger angular separations. Vortex charges of more than 2 are theoretically possible but the peak coupling efficiency is so low that they are impractical. An important point to keep in mind though is that the increased coupling of charge 1 at small separations affects all points, not just the planet. Thus, charge 1 is more sensitive to tip/tilt jitter and the finite angular size of the star than charge 2. A detailed analysis of all the related trades and considerations between the two vortex charges and how they affect the stellar leakage can be found in Ruane et al.\cite{Ruane2019SPIE} 

With the starlight effectively nulled and the planet light transmitted through the fiber, the output of the SMF can be connected to a spectrograph for analysis. Unlike typical nulling interferometry instruments, the VFN coupling efficiency is circularly symmetric, as shown in the inset of Fig.~\ref{fig:MotherFig}c. This means there is no need to modulate the planet signal. Instead, the planet is spectrally detected using template matching or cross-correlation methods which can detect the planet with relatively low signal signal-to-noise requirements per spectral channel\cite{Wang2017,Snellen2015_HDC}. These techniques serve to further suppress the star signal as well as analyze the planet signal to search for key molecules in the planet's atmosphere such as water, methane, and carbon dioxide.

\section{LABORATORY DEMONSTRATION}
\label{sec:Lab_Demos}
\begin{figure}[t!]
    \centering
    \includegraphics[width=0.9\linewidth]{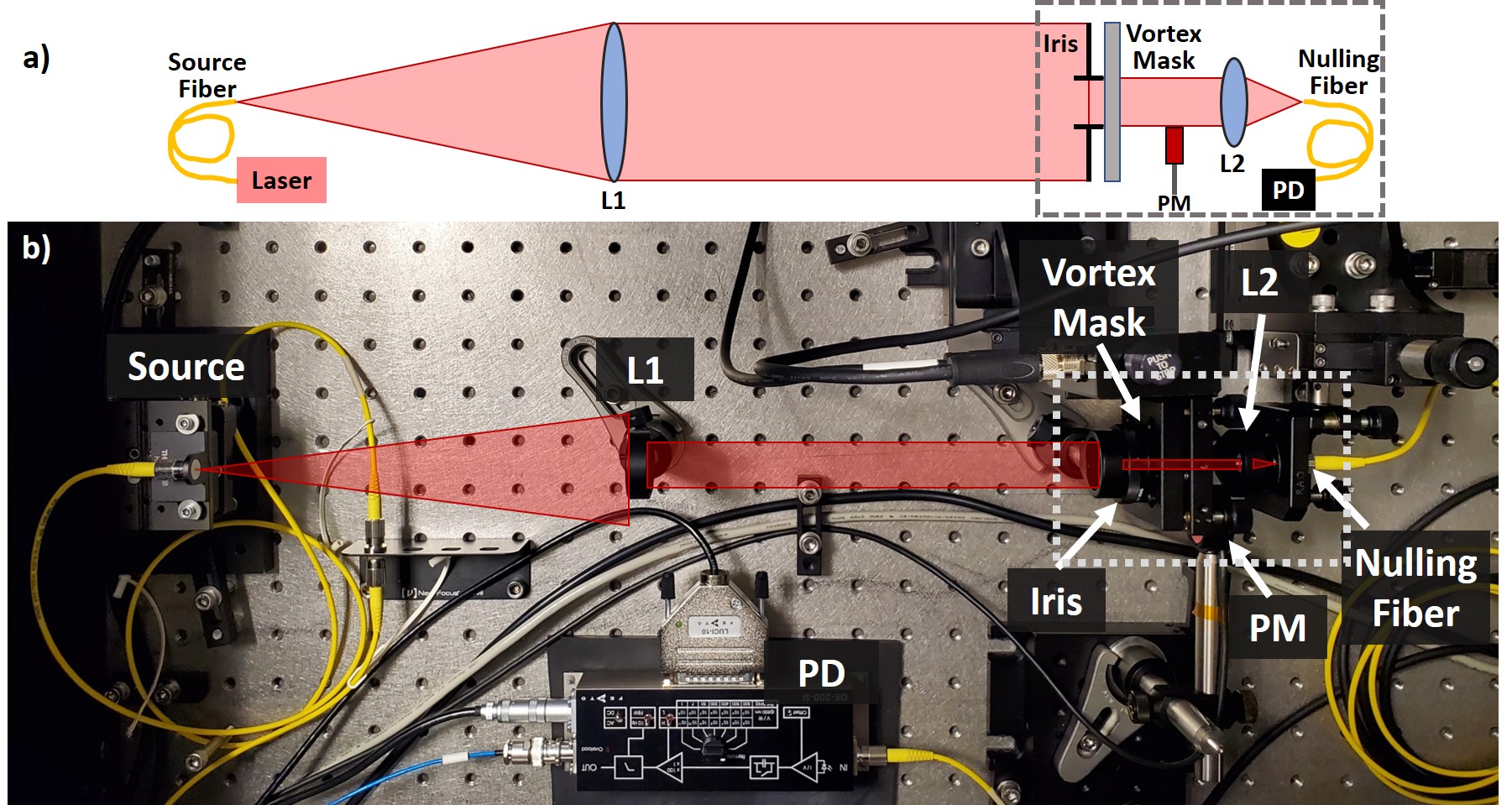}
    \caption{(a) Schematic of the transmissive VFN testbed at Caltech. The source fiber projects light onto the collimating lens, L1. An iris then sets the pupil diameter before passing the beam to the vortex mask. An aspheric focusing lens, L2, images the beam onto the nulling fiber which is connected to a photodiode, PD, for coupling measurements. A retractable power meter, PM, can be moved into the beam path to measure the power for normalization. (b) Picture of the Caltech transmissive VFN testbed.}
    \label{fig:TransmissiveBench}
\end{figure}

In order to test VFN in the lab, we designed and built a dedicated testbed in the Exoplanet Technology Lab (ET Lab) at Caltech. This testbed, shown in Fig.~\ref{fig:TransmissiveBench}, implements a minimalist design with transmissive optics. The light from a fixed SMF acts as the source and is projected onto a long-focal-length lens, L1, so that the fiber tip remains unresolved. The collimated light is then passed through an iris that sets the desired pupil diameter before proceeding to the vortex mask and finally being focused onto the nulling SMF using an aspheric lens, L2. In order to simulate multiple point sources, including the star and planets at various locations, we scan the nulling fiber in the imaging plane. This takes advantage of the fact that the system is shift-invariant such that moving the nulling fiber is equivalent to moving the source fiber. Since the nulling fiber is already actuated in order to accurately place it in the focal plane, this reduces the complexity of the system and avoids having to add more optics to create a second point source for the planet close enough to the star. 

A coupling efficiency measurement requires knowledge of the total light incident on the focal plane. In this transmissive testbed, the incident light is measured using a power meter (PM), that slides into the beam at the end of a scan. The light transmitted by the fiber is then measured throughout the scan on a fiber-coupled photodiode (PD) and those power values are normalized by the PM reading to provide the coupling efficiency. This also leads to detailed 2D coupling maps like the inset from Fig.~\ref{fig:MotherFig}c and as shown in practice in Fig.~\ref{fig:MonoUpdate}a.

\subsection{Monochromatic Experiments}
\label{sec:mono_exp}
We first demonstrated VFN in a laboratory environment in 2019\cite{Echeverri2019_VFN}. At the time, we reported null depths of $6\times10^{-5}$ and a peak planet coupling of 10\% with a charge 1 vortex mask in monochromatic light using a 635~nm laser. The null depth was was a nice match to the expected value of $5\times10^{-5}$ based on the wavefront errors in the system. The planet coupling peak was less than the theoretical maximum of 19\% for a charge 1 vortex but this was due to a mismatched F\# in the system; given the actual F\#, we could only achieve a theoretical max of 12\%. As such, though the coupling efficiency was not as high as can be expected for a vortex fiber nuller, the system and its performance were well understood and we were able to validate much of the theory behind VFN.

\begin{figure}[t!]
    \centering
    \includegraphics[width=\linewidth]{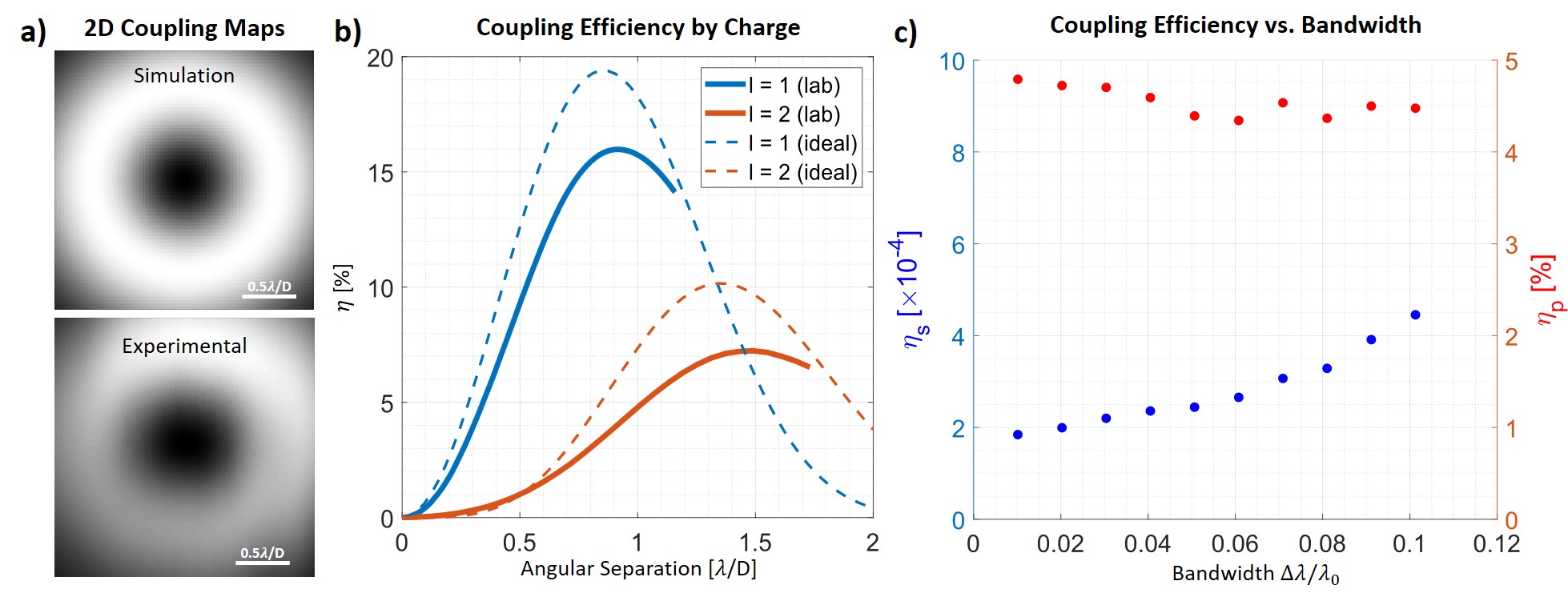}
    \caption{(a) 2D coupling maps with a charge 1 vortex mask in simulation (top) and experiment (bottom). (b) Monochromatic coupling efficiency line profiles for charge 1 (blue) and charge 2 (orange). The ideal coupling, determined from simulation, is shown in the dotted lines while the experimental coupling is shown in the solid lines. (c) Polychromatic experimental coupling efficiency of a charge 2 vortex mask at various bandwidths centered around 790~nm. Points in blue represent the null depth, hence star coupling efficiency, while points in red represent the peak planet coupling.}
    \label{fig:MonoUpdate}
\end{figure}

Since then, the system has been improved and optimized to achieve better performance. The biggest change was that the pupil diameter was adjusted to set an F\# closer that the one required by the SMF core. This, along with a few other minor modifications, resulted in a similar null depth but much higher planet coupling. The new star coupling was measured to be $6.6\times10^{-5}$ which matches the fact that the wavefront error did not change much from when the previous result was reported so we were still limited by the coma aberrations in the system. The azimuthally averaged peak planet coupling was measured at 16\%, putting it much closer to the 19\% theoretical maximum. The full coupling efficiency line profile is shown in the solid blue line of Fig.~\ref{fig:MonoUpdate}b. The coupling still doesn't reach the full potential of an ideal VFN, shown as the dashed blue line in the same figure, so we are tracking down what is causing the discrepancy. We believe this may be due to the calibration between the normalization power meter and the final photodiode so we plan on changing the way this normalization is done. The new method is explained in Sec.~\ref{sec:PoRT}. The reduced coupling could also be due to an uncertainty on the true core diameter. We have set the pupil size based on the manufacturer's specification of the SMF core but there is a large tolerance on this value. Therefore, we also plan on empirically determining the ideal pupil diameter and then repeating these experiments.

We have also tested a charge 2 vortex in the lab. This vortex is a broadband mask designed for $420\text{-}870$~nm light\cite{Tabiryan2017}. We initially tested the vortex using the same 635~nm laser as in the last experiments but found that the performance was not as good as expected. Once we switched to a 780~nm laser, the performance was much better. In this layout, the star coupling was measured at $4.7\times10^{-5}$ while the azimuthally averaged peak planet coupling was 7.2\%. The full coupling profile and the corresponding theoretical coupling are shown in orange in Fig.~\ref{fig:MonoUpdate}b. As with the charge 1 experiments, the coupling is close to but not quite as high as the theoretical maximum. We believe this may be due to the same reasons mentioned before and will check once the charge 1 coupling has been improved.

\subsection{Polychromatic Experiments}
\label{sec:poly_sec}
With the charge 2 mask validated in monochromatic laser light, we tested it in polychromatic light to demonstrate the broadband behaviour of VFN. This was done using a supercontinuum white light source (NKT Photonics SuperK EXTREME) and tunable filter (NKT Photonics SuperK VARIA) to programmatically vary the wavelength and bandwidth of the source. As mentioned in a previous proceeding\cite{Echeverri2019b_VFN}, the system performed well. A null of $4.5\times10^{-4}$ with a simultaneous peak planet coupling of 4.5\% was achieved at 10\% bandwidth centered around 790~nm. Figure~\ref{fig:MonoUpdate}c shows that the peak coupling remained at just under 5\% from 3 to 10\% bandwidth while the null degraded slightly. We plan to repeat these experiments with the upgraded, polychromatic testbed described in Sec.~\ref{sec:PoRT} which is better suited for broadband experiments. 

\begin{figure}[t!]
    \centering
    \includegraphics[width=0.9\linewidth]{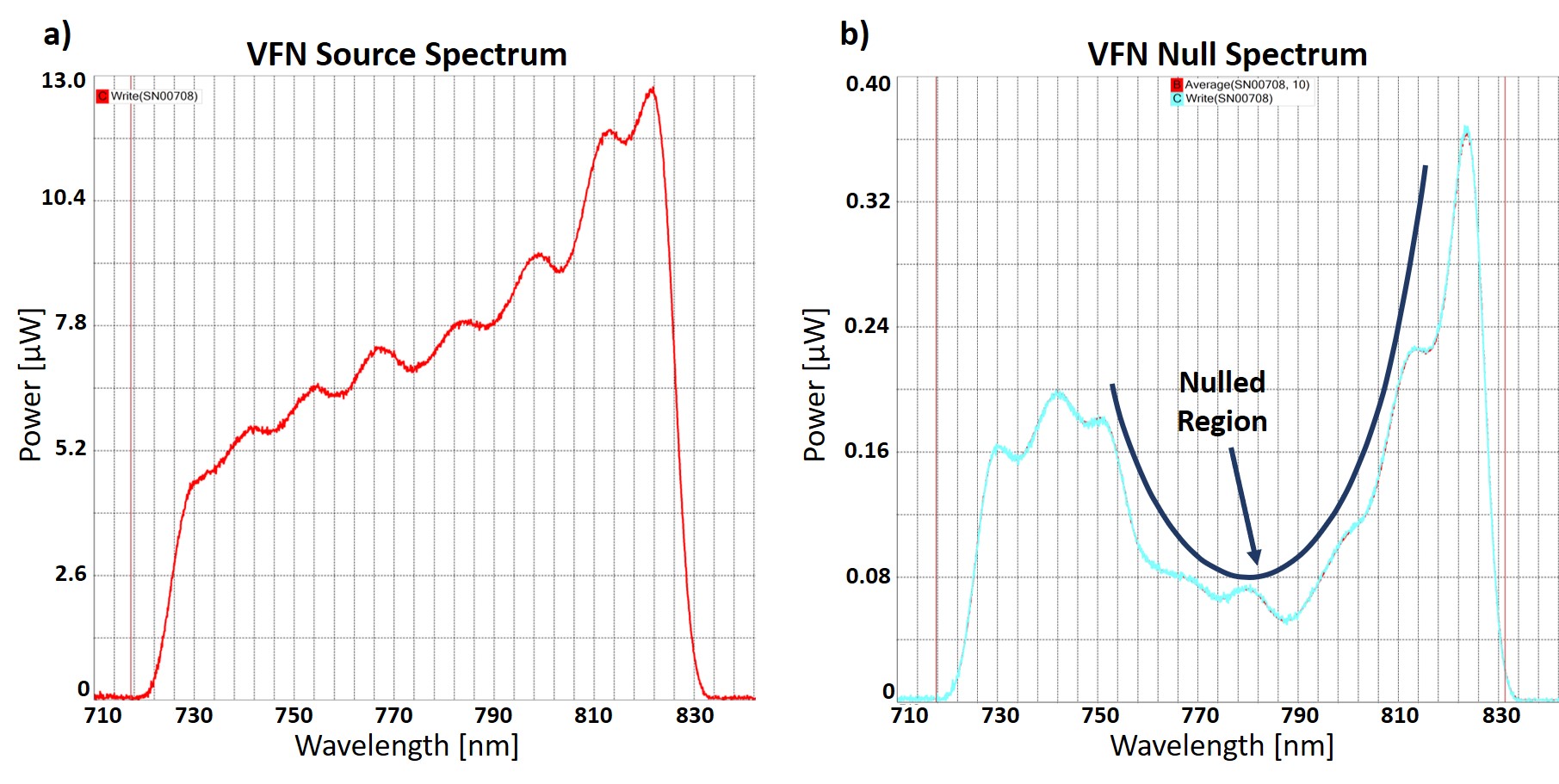}
    \caption{Spectrum of a VFN system as measured on an OSA. A 15\% bandwidth light source centered at 775~nm was used for these measurements. Plot (a) shows the source spectrum as measured before entering the VFN system. (b) Shows the output spectrum after passing through the vortex and nulling fiber. The two spectra were taken under different source power settings such that a direct quantitative comparison is not possible. However, a qualitative comparison can be considered: the output spectrum has the same general underlying form as the input spectrum but there is a clear and deep nulled region where the VFN effect modulated and suppressed the input spectrum.}
    \label{fig:OSASpectrum}
\end{figure}

Another way to analyze the broadband performance of the system is using an Optical Spectrum Analyzer (OSA). For these tests, we replaced the photodiode with an OSA (Thorlabs OSA202C) which provides the full system spectrum from 600 to 1700~nm in a single measurement. Figure~\ref{fig:OSASpectrum}a shows the input spectrum as measured directly from the source fiber while \ref{fig:OSASpectrum}b shows the VFN spectrum measured at the output of the nulling fiber with the fiber aligned to the source to simulate the star's nulled spectrum. The source spectrum has a generally increasing power from 730~nm to 822~nm reflecting the $\sim15\%$ bandwidth window centered at 775~nm that was used for these measurements. The nulled spectrum however has a clear and deep parabola carved out where the the vortex has successfully prevented part of the starlight from coupling into the fiber, thereby modulating the input spectrum. This parabola is centered around 775~nm as expected since that is the wavelength used when optimizing the fiber-vortex alignment to achieve the best null. 

The y-axes of these plots, representing the power in the spectrum, should be considered independently of each other. The two measurements were made under different conditions such that the values are not calibrated to each other. The input spectrum was measured at the source fiber before taking any losses from the system and with the light source at 15\% power to avoid saturating the OSA's detector. The output spectrum, meanwhile, was measured with the source at 100\% power to provide a detectable signal on the OSA and also has various uncalibrated losses due to the iris and optic throughputs. This means the spectra cannot be \emph{quantitatively} compared to each other without further measurements to compensate for these unknown losses and the difference in source power. Nevertheless, a \emph{qualitative} comparison can be performed to see how the fiber nulling effect has modulated the underlying waveform of the input spectrum. This clear sign of a nulled region is a promising indicator that the system is working as expected. 
In the future, we will report calibrated null depths using the OSA. The goal is to determine how the VFN null evolves across large spectral bands in a similar but more direct way to what was done in Fig.~\ref{fig:MonoUpdate}c, including obtaining the coupling efficiency as a function of wavelength across the full band.

\section{UPGRADED TESTBED}
\label{sec:PoRT}
\begin{figure}[t!]
    \centering
    \includegraphics[width=0.9\linewidth]{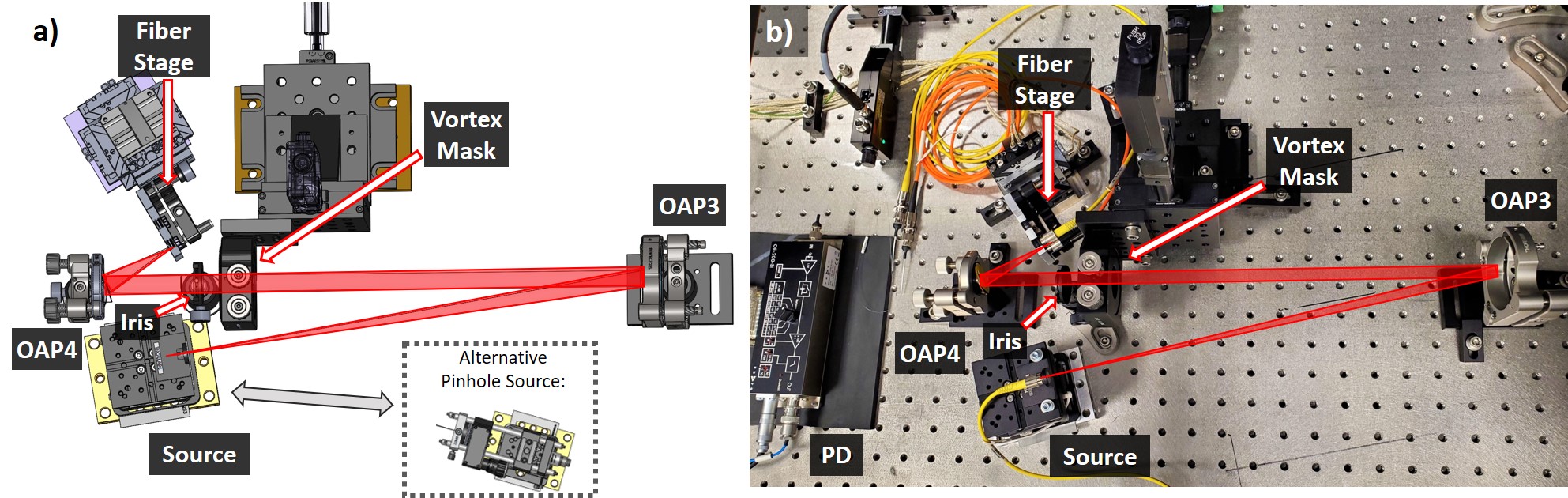}
    \caption{(a) CAD model of the PoRT testbed at Caltech. This design is very similar to the previous transmissive bench but the lenses have been replaced with OAPs to create an achromatic system. The fiber stage has also been upgraded to use high accuracy, closed-loop control actuators and now holds the nulling fiber as well as a multi-mode fiber for normalization. An alternative source which uses a pinhole to ensure the source remains unresolved is shown in the lower right box. (b) Photo of the testbed with all the parts roughly in place prior to the optical alignment. The two fibers are seen mounted on the fiber stage; the yellow fiber is the SMF and the orange fiber is the MMF.}
    \label{fig:HalfPoRT}
\end{figure}

The transmissive testbed used in all the prior experiments has worked well but is limited by the chromatic effects introduced by the lenses, as well as a few other design issues. This drove us to design a new testbed building on the successes of the last but with reflective elements where possible to enable truly polychromatic performance. Figure~\ref{fig:HalfPoRT}a shows a CAD model of the testbed which we have called the Polychromatic Reflective Testbed, or ``PoRT". As before, a single fixed fiber acts as the source and illuminates the collimating off-axis parabolic mirror (OAP), OAP3, that reflects the beam towards the vortex and pupil-defining iris. OAP4, then focuses the light onto the fiber stage. 

The fiber stage has been upgraded from the previous design to hold up to four optical fibers. For now, two fibers will be used: the usual single-mode nulling fiber and a new multi-mode fiber (MMF). The MMF has a core diameter of $105~\mu m$ so that it collects the majority of the incident light. A bifurcated fiber is then used to combine the SMF and MMF inside a single fiber cladding that can be fed into a photodiode. This two-fiber design will enable us to perform the coupling and normalization measurements on the same detector, thereby simplifying the analysis and removing any uncertainties regarding the calibration between multiple detectors. The actuators on the source stage have also been replaced with high accuracy, closed-loop, large range linear actuators from Physik Instrumente (PI). These 3 stages (PI Q-545.240) allow us to position the fibers in the focal plane in X, Y and focus down to an accuracy of 6~nm. The closed-loop control ensures that we know exactly where the desired fiber is located and also removes hysteresis effects that we believe were limiting us from automating the null-finding and optimization procedure on the last testbed. 

We also designed an alternative source stage that reimages the source fiber's tip onto a pinhole so that the angular size of the source can be more accurately controlled. As mentioned in Sec.~\ref{sec:intro}, a charge 1 VFN system is sensitive to the finite size of the source. We have chosen OAP3 and the pupil diameter of the system such that the fiber core should be small enough to remain unresolved for all charge 2 experiments. However, with a charge 1 mask and at shorter wavelengths such as 635~nm, the fiber may be resolved so a pinhole of 5-8~$\mu$m will be needed. This pinhole source design also utilizes a cage system that will allow us to place additional optics into the beampath before the pinhole where their wavefront errors will be spatially filtered. As such, we will now be able to add circular polarizes if needed to remove any polarization-dependent stellar leakage without increasing the wavefront error in the system. 

All of the components of PoRT have arrived and we did a fit test of the full system as shown in Fig.~\ref{fig:HalfPoRT}b. Optical alignment will begin soon and we expect to have the first results with this bench in early 2021. We will first replicate the monochromatic experiments done on the previous bench to ensure that the system is working as expected. After that we will continue with the broadband experiments from Sec.~\ref{sec:poly_sec}. Once VFN has been confidently demonstrated in visible wavelengths, we will change to the near-infrared, specifically to K-band which is centered at $2.2~\mu m$ with a cut-on wavelength of $2~\mu m$. We chose K-band since VFN will be tested on-sky with the Keck Planet Imager and Characterizer (KPIC) instrument as will be discussed in Sec.~\ref{sec:KPIC_Kband}. KPIC VFN will operate in K-band so we want to validate and characterize the vortex mask that will be used on-sky. PoRT's reflective elements should make the transition to these longer wavelengths straightforward since there will be no chromatic effects from the lenses. 

\begin{figure}[t!]
    \centering
    \includegraphics[width=0.9\linewidth]{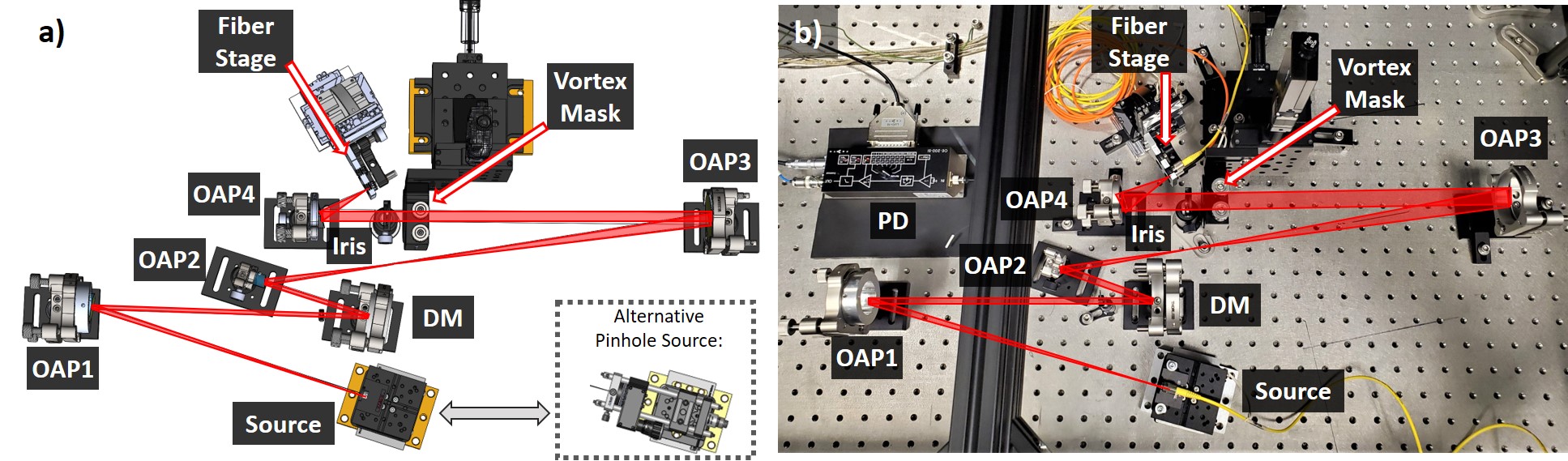}
    \caption{(a) CAD model of the expanded PoRT testbed designed to add a DM to the system. (b) Photo of the expanded testbed after a fit test. The black beam near OAP2 is part of the enclosure that isolates PoRT and is not actually in the beampath.}
    \label{fig:FullPoRT}
\end{figure}

PoRT was designed with an expanded concept in mind. As VFN gets closer to on-sky deployment, we will need to develop wavefront control (WFC) algorithms for it. Since each vortex charge is only sensitive to a specific set of aberrations\cite{Ruane2019SPIE}, we plan on modifying the standard WFC methods used by KPIC to leverage this modal selectivity. By loosening the constraints on some Zernike modes and targeting only the ones that increase the stellar leakage, it should be possible achieve deeper null depths on sky. These WFC development endeavors will require a deformable mirror (DM) to control the wavefront in the system. An expanded version of PoRT has been designed which adds an optical relay to create a second pupil plane for the DM to sit in. Figure~\ref{fig:FullPoRT}a shows the CAD model for this upgraded PoRT layout and Fig.~\ref{fig:FullPoRT}b is a picture of the system after a fit test.

\section{PRELIMINARY K-BAND RESULTS}
\label{sec:KPIC_Kband}
All the demonstrations of VFN in the lab are building towards an on-sky demonstration using the KPIC instrument that is now operational in the Keck II Telescope at the W. M. Keck Observatory in Hawaii\cite{Mawet2016_KPIC,Mawet2017_KPIC,Jovanovic2019SPIE}. KPIC is a phased upgrade to the Keck adaptive optics system and includes two key elements: an infrared pyramid wavefront sensor\cite{Bond2018_PyWFS} and a Fiber Injection Unit (FIU). The FIU adds, among other things, a single mode fiber source to the existing high-resolution spectrograph at Keck. KPIC has recently boasted several rounds of successful on-sky operation and science data has been obtained on several targets of interest\cite{Delorme2020_KPIC}; these results are currently being reduced and analyzed to be published in journal papers soon. 

The next planned upgrade, Phase II\cite{Jovanovic2020_KPIC}, will include several improvements and state-of-the-art sub-modules including a set of Phase Induced Amplitude Apodization (PIAA) optics to increase the fiber coupling efficiency, a high order DM for dedicated WFC, and an atmospheric dispersion compensator (ADC)\cite{Jovanovic2020_KPIC,Wang2020_ADC} to correct for chromatic dispersion in the atmosphere. Phase II will also include a ``coronagraph" sub-module that houses a charge 2 vortex mask thereby enabling VFN on the KPIC. For details on this sub-module, the VFN implementation, and the predicted VFN performance refer to Echeverri et al.\cite{Echeverri2019b_VFN}.

\begin{figure}[t!]
    \centering
    \includegraphics[width=0.8\linewidth]{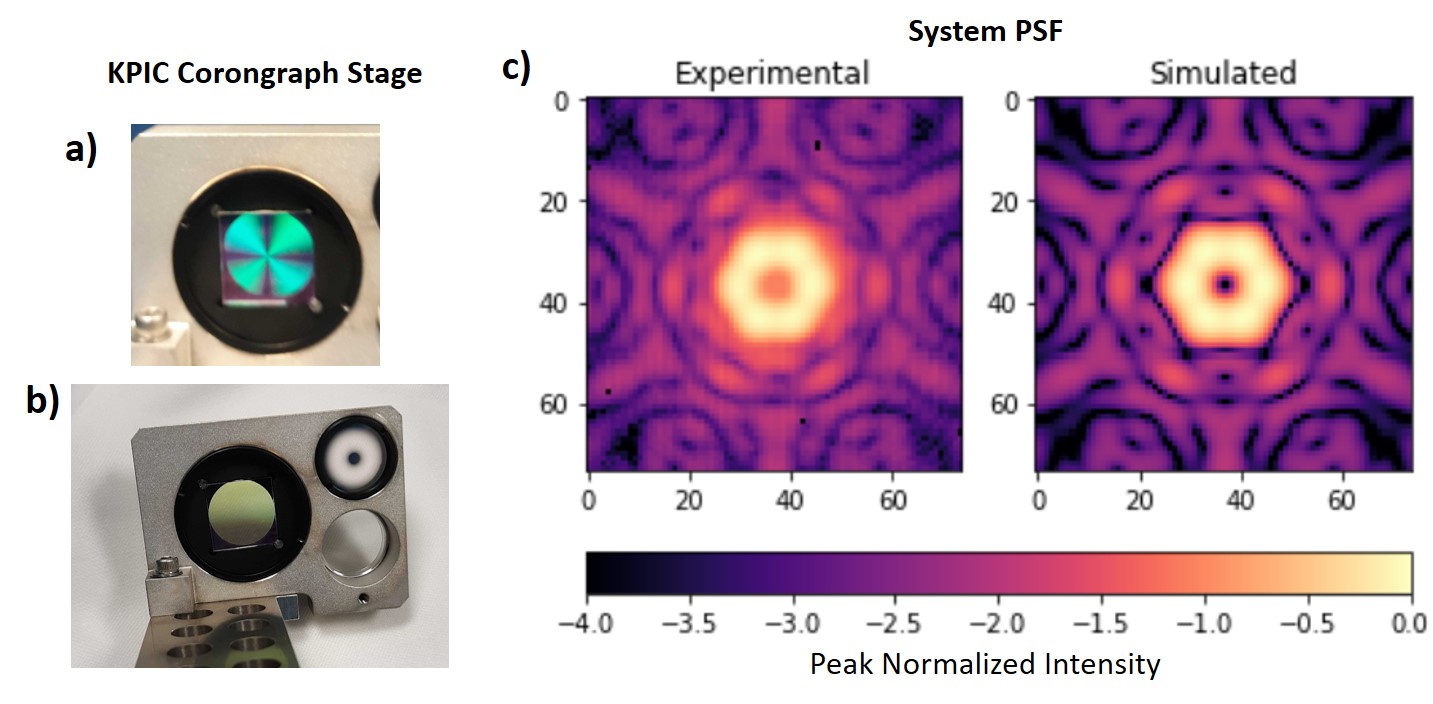}
    \caption{(a) K-band charge 2 vortex glued into its mask holder and seen through crossed polarizers to show the vortex pattern. (b) KPIC ``coronagraph" stage with the apodizer and vortex masks. (c) Experimental vs. simulated PSF for the Keck aperture with the charge 2 vortex mask in the beam.}
    \label{fig:KBand}
\end{figure}

Laboratory tests are being done to validate and characterize the K-band vortex mask that will be used in KPIC. The bulk of these tests will be performed on PoRT prior to integration into the Phase II mechanical plate but since PoRT is not yet aligned, those experiments that can be done without PoRT are now underway. The optical vortex has been bonded to its mask holder and has been mounted in the KPIC ``coronagraph" stage as shown in Fig.~\ref{fig:KBand}a,b. There is currently a near-to-mid infrared testbed in the ET Lab that is used for testing other elements of the KPIC instrument prior to deployment. This bench is relatively simple but it provides a readily-accessible pupil plane where optics can be validated. We used this bench to do some preliminary measurements of the vortex performance. 

The vortex PSF was already imaged last year on this bench and those results were shown in last year's conference. This year we added a Keck-like pupil mask to simulate the hexagonal shape, central obstruction, and spiders present on the Keck aperture. With this mask in place, we retook the images of the vortex PSF. The left plot in Fig.~\ref{fig:KBand}c shows the experimental PSF while the right plot shows the expected PSF based on simulations. There is a close visual agreement between the two but we can quantify the similarity by computing the overlap integral of the convolution of the two images. This computation reveals that the experimental PSF is a 98.6\% match to the simulation.

We then measured the coupling efficiency at the null point as well as at the peak planet location. This system is not ideal for fiber coupling experiments since there is no way way to scan the fiber but there is a flat fold mirror that can be used to move the PSF over the fiber tip. Though the mirror is not at a true pupil plane, it is close enough for a quick and dirty measurement of the coupling. The null point, and hence star coupling, was measured to be just under 1\% while the peak coupling was measured at just over 8\%. The 8\% is relatively close to the theoretical maximum of 11\% for a charge 2 vortex on a Keck-like aperture. It is possible that the loss of 3\% is due to a slight defocus introduced by the vortex mask that cannot be easily fixed with the lack of actuators on this bench. It could also be due to the fact that the 8\% measurement was made by simply tilting the mirror linearly outward from the null in a single direction to scan the coupling line profile. Since the VFN coupling map on the Keck aperture is not rotationally symmetric, it is possible that this line scan passed through one of the regions with slightly lower coupling. The 1\% null depth is much worse than we'd expect from a VFN system with such low wavefront error but given the limitations of this bench and the fact that there is no way to scan the fiber or vortex to optimize the alignment, it's likely that there is room for improvement on PoRT. We believe that once these tests are repeated on PoRT, which is designed for coupling measurements and optimization, we will achieve deeper null depths on the order of what we have demonstrated in visible wavelengths already.

\section{SUMMARY}
\label{sec:Summary}
Building on the extensive heritage of nulling interferometry and fiber nulling specifically, VFN is a new technique for detecting and characterizing exoplanets at small angular separations. By using a vortex mask in the pupil plane, the full telescope aperture can be utilized which in turn improves the system throughput. This design also removes the need to rotate the pupil to modulate the planet signal and instead relies on a spectrograph to extract the planet from behind the star's glare. 

VFN has been demonstrated in the lab at visible wavelengths in both monochromatic and polychromatic light. Laboratory measurements regularly achieve null depths of $<10^{-4}$ with a laser and $<10^{-3}$ in 10\% bandwidth light. The peak planet coupling in the lab has been shown to reach 16\% for a charge 1 vortex and 7\% for charge two. We are now working on maximizing the planet signal to get closer to the theoretical maximum and are planning to use an OSA to measure the full VFN spectrum directly so that the broadband behavior of the system can be evaluated. 

A new testbed has been designed and is being aligned now which will improve the laboratory results. This testbed makes several improvements over the previous transmissive bench:
\vspace{-4mm}
\begin{itemize}
    \setlength\itemsep{-5pt}
    \item OAPs are used to remove chromatic aberrations from the system and improve the broadband performance.
    \item The fiber nulling stage has been upgraded to house 2 fibers so that the nulling and normalization measurements can both be made in the same image plane and on the same photodetector.
    \item High precision, closed-loop actuators on the fiber stage provide better accuracy when scanning the nulling fiber and will also enable automatic optimization scripts that can find the best alignemnt for the system.
    \item An alternative source stage has been designed to use a pinhole to ensure that the source remains unresolved when working with a charge 1 vortex at short wavelengths. 
    \item An expanded version of the testbed also creates a second pupil plane for adding a DM so that WFC experiments can be performed on PoRT.
\end{itemize}
\vspace{-3mm}
Once this testbed has been validated in the visible, we will switch to the near infrared with a K-band source and vortex. The K-band experiments will directly inform the development and deployment of the VFN on-sky demonstration on KPIC.

In preparation for the transition to K-band, we imaged the PSF of the vortex mask that will be used on sky in KPIC. The mask was placed in a near infrared testbed with a Keck-like pupil mask to replicated the Keck aperture. The system PSF was a 98.6\% match to the expected PSF based on simulations. We then took coarse measurements of the star and peak planet coupling efficiencies and got values of 1\% and 8\% respectively. The null depth is not as good as expected but planet coupling is close to the theoretical maximum of 11\%. The current near-infrared bench used for these experiments is not ideal for coupling measurements so we believe that repeating these experiments on PoRT should recover the null depths demonstrated already at visible wavelengths. 

Once VFN has been demonstrated in the near-infrared in the lab, it will be moved to KPIC for an on-sky demonstration. This deployment is part of the Phase II upgrade to KPIC that is scheduled for the end of 2021. The results of that demonstration will inform the design of a possible VFN mode on future instruments such as MODHIS and HISPEC\cite{Mawet2019_HISPEC}. If a VFN mode were implemented on a large space telescope, such as NASA's LUVOIR mission concept\cite{Bolcar2018_LUVOIR}, it would complement the coronagraphic mode to enable the characterization of Earth-like exoplanets at and within the diffraction limit\cite{Ruane2018_VFN}.

\acknowledgments 
Daniel Echeverri is supported by a NASA Future Investigators in NASA Earth and Space Science and Technology (FINESST) fellowship under award \#80NSSC19K1423. This work was supported by the Heising-Simons Foundation through grants \#2019-1312 and \#2015-129. Part of this research was carried out at the Jet Propulsion Laboratory, California Institute of Technology, under a contract with the National Aeronautics and Space Administration (80NM0018D0004). The authors wish to recognize and acknowledge the very significant cultural role and reverence that the summit of Maunakea has always had within the indigenous Hawaiian community. We are most fortunate to have the opportunity to conduct observations from this mountain.


\small
\bibliography{Library} 
\bibliographystyle{spiebib} 

\end{document}